# An Analysis of Triangulation in Geometrically Noisy Environments using Mathematics

Noah M. Kenney

*Abstract*—This paper uses mathematics to analyze the challenges of geometrically noisy environments on triangulation. Given widely accepted algorithmic triangulation methods, such as $O(n \ln n)$ or a simpler $O(n^3)$ method, we can mathematically prove that triangulation of any two dimensional polygonal region is possible, albeit impractical in some cases. Further, we consider the implications of environments in which a z-axis is present, as seen in cellular triangulation. In many of the cases where consideration of the z-axis is necessary, we recognize the absence of a fixed or known point of origin and consider methods of addressing this challenge.

## I. Two-Dimensional Polygonal Triangulation

WHILE any two-dimensional polygonal figure can be triangulated, many polygonal figures can only be triangulated using an infinite number of triangles [1]. Through Meisters' Two Ears Theorem, a diagonal triangulation of polygons with three or more vertices is possible, given that an interior triangle can be formed between the two end vertices (represented as A and C and in a polygon defined by vertices A, B, and C) [2].

A polygon of *n* vertices (*n-gon*), will have (*n-2*) triangles, formed by diagonals of known quantity, calculated as (*n-3*). This results from the ears formed by consecutive vertices, of which there may be no more than three. Thus, any three consecutive vertices will form an ear in a Jordan polygon, which can be represented as $P=V_1V_2V_3V_4\ldots V_nV_1$ [3]. Mathematically, the Proof of Lemma effectively shows that "the degree of each of the nodes is at most three because a triangle in a triangulation may be adjacent to at most three other triangles" [3].

We can easily prove this for polygons with few vertices. For example, a polygon in which *n*=3 (triangle), will have (*n-2*) triangles, or one triangle. The same polygon will have (*n-3*) diagonals, or zero. For a polygon in which n=4 (rectangle), the same calculations apply. The rectangle will result in two triangles formed by one diagonal. In geometrically noisy environments, it may be necessary to triangulate using more than one diagonal, in which case triangulation may prove more complex.

Figure 1 displays four, out of infinite, possible methods of triangulation of a square. First, we see the most simplistic form of triangulation (two triangles formed by one diagonal). Each additional presented square shows increasingly complex forms of triangulation, formed by additional vertices. While a square does not require an infinite number of triangles for triangulation, it is possible to triangulate with no outer bound on the number of triangles formed, assuming that there is no such bound on the number of diagonals. Further, triangulation of a square may not only include right triangles, but also acute scalene triangles, equilateral triangles, isosceles triangles, and obtuse scalene triangles [4], as seen in the third and fourth squares presented in Figure 1 (below).

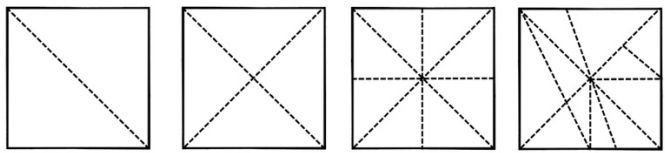

Fig. 1. Triangulation of squares.

The same principles hold when dealing with triangulation of parallelograms, as seen in Figure 2 (below).

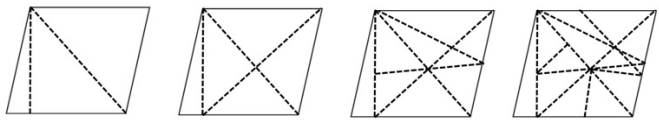

Fig. 2. Triangulation of parallelograms.

While the triangulation of squares and parallelograms can largely be accomplished by bisecting opposite vertices of the polygon, triangulation of trapezoids cannot be accomplished in the same way, given the lack of horizontal symmetry. Regardless, as is the case with both squares and parallelograms, triangulation can be accomplished with only one diagonal seen in Figure 3 (below).

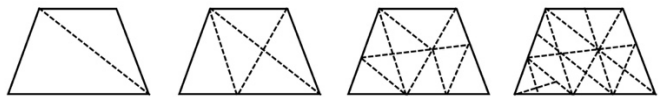

Fig. 3. Triangulation of trapezoids.

It is reasonable to consider the reason why triangulation may be necessary, or at least helpful. Namely, in a figure of three or more vertices, such that points A, B, and C all exist in different coordinates on a two-dimensional plane, the distance of line segment AB plus the distance of line segment BC will be longer than the distance of the hypotenuse represented by side AC.

Thus, the distance between points for the purposes of calculations may be reduced through the use of triangulation, as proved by the Pythagorean Theorem [5].

## II. Two-Dimensional Circular Triangulation

Proving that triangulation of polygonal figures is always possible in any two-dimensional plane is fairly simplistic mathematically, although triangulation itself may be tedious. In contrast, triangulation of a circle is significantly more complex, and requires many more inherent assumptions in the definition of circular triangulation, which is "a closed subset of the closed disc whose complement is a disjoint union of open triangles with vertices on the circumference of the circle" [6].

True triangulation within a circle would require an infinite number of touching vertices, covering the entire circumference of the circle. This may prove impossible in a circle of unknown origin or size. Regardless, an approximation of triangulation can be accomplished with as few as three vertices, each located on the circumference of the circle. In this case, we can ignore arc length and focus solely on the polygonal region created by the vertices. Thus, in circular triangulation approximations, the same principles of two-dimensional polygonal triangulation apply. This is ideal, as two-dimensional triangulation is the most simple and fastest form of triangulation. A simplistic view of a three sided polygon (triangle) created from a circle with three vertices located on the circumference is found in Figure 4 (below). As seen in the example, with three vertices, the accuracy of triangulation is low, as seen in the area of the space between the outer border of the triangle and the circumference of the circle as opposed to the area of the triangle itself.

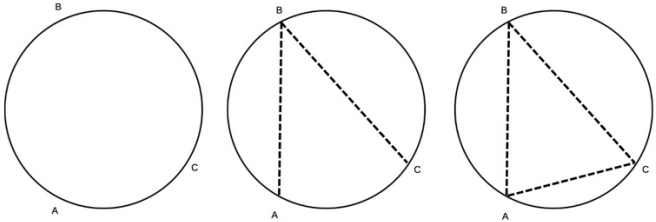

Fig. 4. Triangulation of circles.

The accuracy of triangulation increases with each additional point added around the circumference of the circle. However, it is worth noting that the effects of this principle are diminishing, with the greatest increase in accuracy found in the shift from three vertices to four vertices.

## III. Three-Dimensional Triangulation

Three-dimensional triangulation of straight-sided shapes follows many of the same principles of two-dimensional polygonal triangulation. In fact, a standard three-dimensional plane (seen in Figure 5) can be comprised of two x-y coordinate planes, in which one of the planes is inverted and rotated 180 degrees around the x-axis.

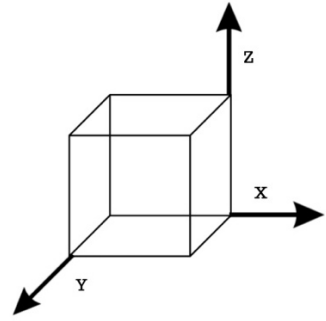

Fig. 5. Diagram of three-dimensional plane (X, Y, and Z).

Translating a three-dimensional object into two-dimensional planes can perhaps best be represented by a cube. The process requires two steps. First, a cross-section of the three-dimensional objects must be produced, which will result in the creation of two congruent triangular prisms (cross-sections). This step is observed in Figure 6 (below).

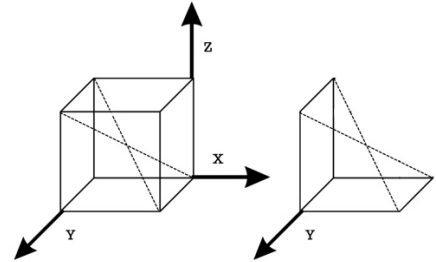

Fig. 6. Diagram of cross-sectional cube (left), and the resulting triangular prism (right).

Second, we can recognize that the resulting cross-section is comprised entirely of two-dimensional polygons. First, we see two triangular sides, thus requiring no triangulation. Second, we see two parallelograms, each of which is easily triangulated as seen previously in this paper. Additionally, we see the five vertex polygonal region created by the cross section, which can be divided into two distinct triangles. These components are observed in Figure 7 (below).

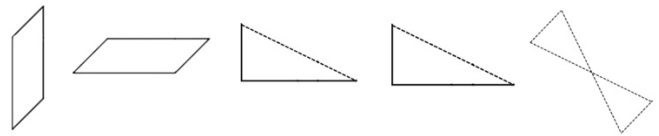

Fig. 7. Diagram of the two-dimensional components of a cube cross section.

Thus, the transition of the three-dimensional cube to six two-dimensional polygons is complete and triangulation is now possible. While the same principles apply for other straight-sided three dimensional objects, the translation of the three-dimensional object to two-dimensional polygons may be more complex. Further, in geometrically noisy environments, it may be possible to triangulate using a number of polygons which may be in conflict with efficiency. An example of this possibility is seen in Figure 8 (below), in which case there are four distinct cross-sections, each comprised of a four vertex figure, instead of the more simplistic approach of one cross-section of five vertices as seen previously in Figure 6.

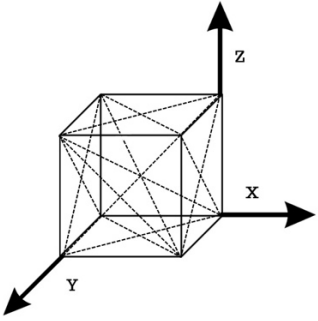

Fig. 8. Diagram of square with multiple cross sections connected at vertices.

In cases where the three-dimensional object is comprised of multiple smaller three-dimensional objects, it may be possible to first break the object into smaller three-dimensional objects before producing the cross-sections.

## IV. THE ADDITION OF GEOMETRIC NOISE

Geometric noise may result in the necessity for more complex triangulation. Noise may prevent the bisecting of two dimensional polygonal regions, thus causing the number of vertices, $n$, to increase as the number of diagonals, $d$, increases. Further, triangulation of the point in which the noise originated may prove impossible, although it may be possible to detect the noise mathematically using triangulation [1].

As an example, in the case of cellular triangulation, data signals can generally be transmitted off of a minimum of three cell towers, though the triangulation with noise may require more than ($n-3$) diagonals, and thus more than ($n-2$) triangles. This can be seen in Figure 9 (below).

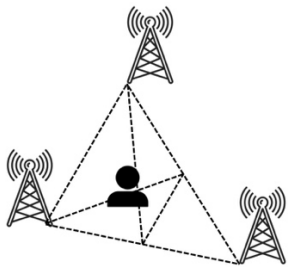

Fig. 9. Cellular triangulation with three towers.

If the cellular user is located within a geographically noisy environment, triangulation may prove easier with four or more cellular towers, as this allows for reasonably accurate two-dimensional polygonal point of origin estimations. This can determine an estimation of the cellular user's physical location, allowing for a second round of triangulation using three cellular towers, as seen previously in Figure 9. In this case, the first round of triangulation may stem from cellular towers of $n$ vertices, where each vertex represents a cellular tower and where $n$ is greater than or equal to four. A model of triangulation using four cellular towers is shown in Figure 10 (below).

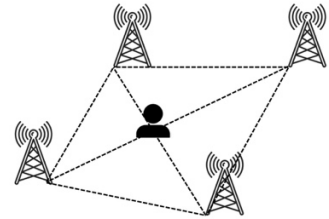

Fig. 10. Cellular triangulation with four towers.

The initial accuracy of triangulation will depend on a variety of factors. One influential factor is the position of the cellular user in reference to the point of intersection between the two triangles aligned on the central axis of the polygonal plane created between four or more cellular towers. For example, in Figure 11 (below), the triangulation will take significantly longer and ultimately result in the creation of more triangles, given that the user is more distant from the point of origin.

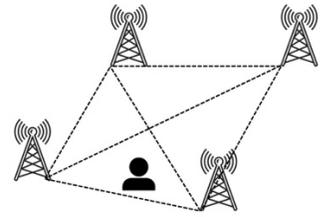

Fig. 11. Cellular triangulation with four towers, when the cellular user is distant from point of intersection.

While this analysis is done in the second dimension, we can assume that cellular triangulation will occur in a three-dimensional environment. Thus, as previously discussed, it will be required to split the three-dimensional environment into standard three-dimensional objects, produce the relevant cross-sections, and then translate the cross sections into two-dimensional polygonal figures. By creating a cross section of the three-dimensional object, the surface area of the polygonal region can be calculated and geographically mapped, as is necessary in many cases of cellular triangulation.

When the analysis is done in the third dimension, we are faced with an additional challenge of not knowing the z-axis point of origin, due to fluctuating land heights relative to sea level. In these cases, triangular approximations in the second dimension are more practical and efficient.